**Title**

Atomic observation on diamond (001) surfaces with non-contact atomic force microscopy

**Authors**

Runnan Zhang[1], Yuuki Yasui[1], Masahiro Fukuda[2], Taisuke Ozaki[2], Masahiko Ogura[3], Toshiharu Makino[3], Daisuke Takeuchi[3], Yoshiaki Sugimoto[1†]

**Affiliations**

[1] Department of Advanced Materials Science, The University of Tokyo, Kashiwa, Chiba 277-8561, Japan.

[2] Institute for Solid State Physics, The University of Tokyo, Kashiwa 277-8581, Japan.

[3] Advanced Power Electronics Research Center, National Institute of Advanced Industrial Science and Technology (AIST), 1-1-1 Umezono, Tsukuba 305-8568, Japan

[†] ysugimoto@k.u-tokyo.ac.jp

**Abstract**

To achieve atomic-level characterization of the diamond (001) surface, persistent efforts have been made over the past few decades. The motivation behind the pursuit extends beyond investigating surface defects and adsorbates; it also involves unraveling the mystery of the smooth growth of diamond. However, the inherently low conductivity and the short C-C bonds render atomic resolution imaging exceptionally challenging. Here, we successfully overcame these challenges by employing non-contact atomic force microscopy with reactive Si tips. Atomic resolution imaging was achieved even at room temperature. With density-functional-theory calculations, we clarified that the critical factors for atomic resolution are in the formation of tilted C-Si bonds between scanning probes and surfaces, along with reordering of the surface C-C dimers. Implications of the findings extend beyond the realm of surface characterization. The present atomic-resolution microscopies drive future advancements in diamond technologies by providing avenues for identifying dopants and constructing artificial nanostructures.

**Introduction**

The wide-gap semiconductor properties of diamond facilitate the development of post-silicon semiconductor devices such as metal-oxide-semiconductor field-effect transistors (1) and quantum sensors with nitrogen center vacancies (2, 3). Diamond surfaces are also employed for heterostructures with silicon (4), graphite (5), and aluminum oxide (6). A successful fabrication of these devices depends on the heterostructure interfaces and dopant configurations. Atomically resolved microscopy plays a crucial role in clarifying the basis of the substrates.

Previous reports have focused on attaining atomic resolution for diamond surfaces. Atomic resolution has been successfully achieved for hydrogenated diamond (001) surfaces through scanning tunneling microscopy (STM) measurements (7-9). The presence of a hydrogen layer renders the surface electrically conductive (10, 11). However, obtaining high-resolution images for clean diamond (001) surfaces is challenging (12-14), due to diamond's large band gap, short C-C bonds, and delocalized one-dimensional electronic states along the dimer rows.

Another approach for surface atomic detection is the non-contact atomic force microscopy (NC-AFM). Silicon, crystallized in the diamond structure, might be one of the most studied materials with the NC-AFM, in which dangling bonds form bonds to a reactive tip. This gives a strong attractive force with the tips, facilitating atomic resolution imaging (15, 16). In contrast, to the best of our knowledge, there is no successful report for sensing single atoms in diamond surfaces (17).

Here, we report on the first observation of individual carbon atoms on diamond (001) surfaces by using NC-AFM. We have elucidated the mechanism of atomic resolution by force curve measurements and density-functional-theory (DFT) calculations. This is highlighted with control measurements on hydrogenated diamond (001) surfaces. DFT calculations provide insights into a reordering of dimer bonds during a tip approach, elucidating the underlying imaging mechanisms. Further analyses of the closest Wannier function (CWF) show a possible orbital for a bond formation to a scanning tip.

**Results**

**Atomic resolution imaging.**

Geometry optimizations based on DFT were performed to analyze the geometrical surface structure. The five-atomic-layer $(2 \times 1)$ surface models with the lattice constant of $5.04 \text{ Å} \times 2.52 \text{ Å}$ for hydrogenated diamond (001) and clean diamond (001) were calculated using the DFT (Fig. 1). The bottom two atomic layers are fixed during the geometry optimizations. Dimerization occurred on both surface models. On the hydrogenated diamond (001) surface, the length of hydrogen dimers is $L_\text{H}^\text{model} = 2.47$ Å, while on the clean diamond (001) surface, the length of carbon dimers is $L_\text{C}^\text{model} = 1.39$ Å. These calculation results align with a previous report (18). It is worth noting that $L_\text{C}^\text{model}$ is very close to the bond length in benzene (C-C in ethane 1.54 Å, graphene 1.42 Å, benzene 1.39 Å, and C=C in ethylene 1.34 Å).

A chemical structure imaging of single molecules was demonstrated, where CO molecules are picked up with scanning tips to prepare well-defined and sharp tips (19). The same method was applied to graphene (20) and $C_{60}$ (21). Because CO molecules hardly form bonds to sample atoms, the imaging is maintained in the repulsive-force regime. While this method allows bond imaging, it is difficult to visualize individual carbon atoms. When a CO-functionalized tip is used for clean diamond surfaces, carbon atoms are not resolved as individual corrugations (refer to Fig. S1 for a simulated AFM image). This exemplifies a limitation in the use of CO functionalized tips. Therefore, we used reactive Si tips to image clean diamond surfaces with attractive chemical bonding forces.

Figure 2(A) displays a NC-AFM topographic image of a hydrogenated diamond (001) surface. The $(2 \times 1)$ reconstruction is manifested by the deeper trench between the dimers. Individual hydrogen atoms are observed as in a previous NC-AFM study (17). The black arrow refers to the hydrogen dimers, and its length is denoted as $L_H^{AFM}$. We obtained $L_H^{AFM} = 2.45 \pm 0.18$ Å by averaging from multiple images. This measured value agrees well with the $L_H^{model}$ of 2.47 Å.

A NC-AFM image of a clean diamond (001) in Fig. 2(B) reveals individual carbon atoms, with the $(2 \times 1)$ surface reconstruction clearly visible. We employed the technique of the active imaging, for which a reactive Si tip is operated near the maximum attractive interaction regime at a close distance. This approach demonstrated a successful imaging of silicene (22). A true atomic resolution allows atomic-level analysis of defects. A single dimer vacancy called Type A defect and double dimer vacancies called Type B defect are observed as in Figs. 2(C) and (D). Besides, we did not observe any single carbon vacancy and a pair of missing carbon atoms along a dimer row, called Type C defect, which can be observed on Si(001) surfaces. Previous calculations on diamond predicted that single-atom vacancies on surfaces are difficult to diffuse into the bulk and tend to exist as dimer defects (Type A) (23). Our calculations show that the formation energy of Type C is much higher than that of Type A (refer to Fig. S2), which is also supported by the present NC-AFM observations.

The individual atoms appear much smaller on the diamond (001) surface than on the Si (001) surface, as shown in Fig. S3. The spatial localization of the carbon orbital contributes to the high spatial resolution, which was suggested as light-atom probes (24). In addition, it is important to spatially separate the force signals of individual carbon atoms for high resolution. The dimer length, denoted as $L_C^{AFM}$ in AFM images, is found to be $2.17 \pm 0.22$ Å from several images. This value is significantly larger than the model dimer length $L_C^{model}$ of 1.39 Å. The apparently longer atomic spacing is also important for the high resolution. The mechanism for AFM imaging contrasts, C-C bond reconstructions by the tip proximity, and force fields detected in AFM will be discussed in the following sections to explain the high resolution.

**Force spectroscopy to reveal AFM imaging contrasts.**

We employed the force-spectroscopy method to quantitatively evaluate the interaction force between the tip and the surface atoms. Figures 3(A) and (B) show the force spectrum measured above apparent atomic positions on the hydrogenated diamond (001) and the clean diamond (001), respectively. The short-range force curves were acquired by fitting and subtracting the long-range force from the frequency-versus-tip-surface distance curves (refer to Fig. S4). The noticeable contrast in the maximum attractive force between the two surfaces indicates that bond formation occurs solely on the clean surface. Specifically, the maximum attractive forces are $-0.3$ nN for the hydrogenated diamond (001) and $-2.7$ nN for the clean diamond (001). DFT calculations are used to rationalize the experimental force curves. The maximum attractive force is found to be $-0.29$ nN for the hydrogenated diamond (001) and $-3.35$ nN for the clean diamond (001). The experimental and theoretical curves exhibit good agreement. These values imply the formation of a strong Si-C chemical bond on the clean diamond surface, while a weak physical interaction dominates on the hydrogenated surface. Hydrogen-terminated carbon atoms are inert and do not form chemical bonds to a reactive Si tip, as inferred from hydrogen-terminated Si(111)-($7 \times 7$) surfaces (25). The strong attractive force in the clean surface accompanies a dissipation signal (Fig. 3(D)), suggesting shifts in the atomic position at the tip approach (26, 27). Such dissipation signal is absent in the hydrogenated surface as in Fig. 3(C).

**Bond modification with tip approach.**

DFT calculations for the clean diamond (001) reveal a clear relationship between the tip-surface distance and the dimer length, elucidating the bond re-ordering, as depicted in Fig. 3(F). When the tip is positioned away from the surface, $L_C^{DFT}$ remains constant at 1.39 Å. However, as the tip approaches the surface, it rapidly increases to 1.55 Å, indicating the transformation of the benzene-like bond into a single-bond-like state. In contrast, on the hydrogenated diamond (001), $L_H^{DFT}$ exhibits negligible variation with the tip-surface distance, as illustrated in Fig. 3(E).

To further clarify the observed apparent dimer length in the NC-AFM image (Fig. 2(B)), we performed additional numerical calculations, as depicted in Fig. 4(A). In this simulation, the tip moves horizontally across the dimer at the height corresponding to the maximum attractive force. Figure 4(B) shows displacement of dimerized carbon atoms as a function of the lateral tip position $x$. For $x < -2$ Å, both carbon atoms shift to the left due to the attractive force exerted by the tip. At $x = -2$ Å, a Si-C bond initiates between the tip and the C atom on the left, prompting the latter's continued movement to the left. Concurrently, the C atom on the right starts moving rightward until $x = -0.4$ Å, signifying an increase in the dimer length (Fig. S5). As the tip approaches near $x = 0$, the Si-C bond breaks, decreasing the dimer length. The tip's continued rightward motion leads to the C atom on the left shifting leftward again, as a new Si-C bond forms with the C atom on the right, causing a rise in the dimer length. For $x > 0.6$ Å, the C atom on the left also begins moving rightward under the tip's attractive force. In summary, as the tip moves across the dimer, the dimer length that the tip detects can be elongated at most 0.16 Å (indicated by the orange dotted lines in Fig. 4(B)). Consequently, this contribution alone may not adequately account for the apparently longer dimers observed in the topographic images. To explain the experiment, it is necessary to investigate an AFM observable, namely the force field.

**Force field analysis**.

When a bond between a tip and a sample atom deviates from the vertical direction, the tip experiences a substantial force directed away from the top of the atom. This phenomenon contributes to a shift in AFM topographic images, given that the constant-frequency-shift feedback approximately follows the constant force contour. The green curve in Fig. 4(C) shows the simulated force as a function of the lateral tip position. We observed the minima in the attractive force position at $x = 1.09$ Å and $x = -1.01$ Å. Consequently, an apparent dimer length is determined to be $L_C^{DFT} = 2.10$ Å. Hence, the tilt of the Si-C bond is identified as the primary factor for the increase in $L_C^{AFM}$ in the AFM topographic images. Note that this value encompasses the contribution of the bond modification as well.

The bond angles are simulated using the CWF simulation (28). For each C atom, four CWFs are formed, and three of them are almost fully occupied, leading to the formation of the dimer structure. The remaining CWF extends towards the vacuum, as illustrated in Fig. 4(A), poised to establish a bond. We infer that the CWF results in the chemical bond to a tilted orientation, and thus the maximum attraction occurs away from the top of atoms.

It is known that a flexible tip leads to an apparent increase in carbon bonds in NC-AFM images as discussed in $C_{60}$ with CO tips (21). To investigate the effect of tip flexibility on the present findings, we conducted calculations for a tip with atomic relaxation, as shown in the blue curve in Fig. 4(C). It was observed that the minima are more distinctly separated for a relaxed tip (blue) than for a fixed tip (green). Specifically, the minima in the blue curve occur at $x = 1.15$ Å and $x = -1.24$ Å, corresponding to $L_C^{DFT} = 2.39$ Å. This value is also in the range of the experimental error. It is crucial to note that this additional contribution is not essential for elucidating the current AFM observations. Furthermore, the movement of the silicon atom at the probe apex is detailed in Fig. S6.

**Discussion**.

Figure 4(D) presents a summary of the aforementioned contributions. To statistically determine the $L_H^{AFM}$ and $L_C^{AFM}$, multiple high-resolution NC-AFM images were analyzed. For hydrogenated diamond (001) surfaces, the obtained value for $L_H^{AFM}$ is $2.45 \pm 0.18$ Å (represented in black circle), which agrees with the modeled value of $L_H^{model} = 2.47$ Å (depicted in black triangle). Conversely, for clean diamond (001) surfaces, $L_C^{AFM}$ is measured at $2.17 \pm 0.22$ Å (illustrated in the red circle), exhibiting a notable deviation from the modeled value of $L_C^{model} = 1.39$ Å (indicated in the red triangle). The repositioning of carbon atoms caused by the tip-induced relaxation partially accounts for this discrepancy (depicted in the orange square). Considering the force field emanating from the CWF protruding into the vacuum at a tilted angle, we successfully elucidated the AFM apparent dimer length (represented in the green inverted triangle). The introduction of tip atom relaxation additionally enhances the apparent dimer length (depicted in the blue pentagon). It is noteworthy that the tip relaxation is not essential for explaining the observed phenomena.

In summary, the present NC-AFM technique successfully achieved atomic resolution imaging of carbon atoms on clean diamond (001) surfaces. While the hydrogenated diamond (001) surfaces exhibit inert behavior to the tips, a pronounced contrast was observed on the clean diamond (001) surfaces. The discovery of a strong attractive force and an enhanced dissipation signal on the clean surfaces suggests the formation of bonds between the tip and surface carbon atoms, ultimately leading to a re-ordering of the dimer bonds. The DFT calculations not only reproduce this strong attractive force but predict the breaking of the dimer. Additionally, the formation of the bond is represented with the CWF. The atomically resolved microscopy presented here relies on the formation of bonds between the tips and individual carbon atoms on clean diamond (001) surfaces. This chemical bonding force can be leveraged for chemical identification and atom manipulation, as demonstrated for the potential to identify dopants and construct artificial nanostructures (29, 30). Harnessing the chemical bonding force in this technique holds promise for advancing power devices on diamond surfaces, offering a means to identify dopants and build artificial nanostructures with atomic precision.

**Materials and Methods**

**Sample growth.**

We prepared diamond homoepitaxial thin films using the chemical-vapor-deposited (CVD) technique on a high-pressure high-temperature (HPHT) synthetic (001) single-crystal diamond substrate ($3 \times 3 \times 0.4 \ \text{mm}^3$). In the reaction chamber, hydrogen, methane and trimethylborane gases were used with a total flow rate of 400 sccm and the gas pressure was maintained at 25 Torr. Specific reaction conditions included a $C/H_2$ ratio of 0.3% and a B/C ratio of 0.005%. We heated the base to 800°C, which was continuously maintained for 4 hours, while 750 W of power was supplied in a microwave field (frequency of 2.45 GHz). This step helped to excite the molecules on the surface to the plasma state and induced the growth of the diamond film. Eventually, we grew diamond (001) films of approximately 400 nm thickness on diamond seeds with an in-plane angle of less than 0.1 degrees. Subsequently, we performed a hydrogen etching process for about a few minutes to further optimize the surface properties of the diamond film (31).

**AFM measurements.**

In prior to the AFM measurements, the hydrogenated diamond samples were annealed with SiC heaters in ultra-high vacuum conditions (base pressure of $5 \times 10^{-9}$ Pa) to remove adsorbates and hydrogens. The hydrogenated diamond surfaces were obtained with annealing up to 1040 K for 2.5 hours. The clean diamond (001) surfaces were obtained with further annealing up to 1290 K for 2 hours.

A custom-built AFM based on optical interferometry was operated at room temperature. AFM measurements were performed in an ultra-high vacuum chamber at room temperature. We used silicon cantilevers with resonant frequency $f_0$ of 150-167 kHz, elastic constant $k$ of 28-36 N/m, and Q-factor of around 10000, after cleaning with $Ar^+$ ion sputtering. The oscillation amplitude $A$ was set to 160-200 Å with the operation in the frequency-modulation mode. Sample bias $V_s$ of $-0.55$ V to $1$ V was applied to minimize the electrostatic forces.

For the force curves, an atom-tracking technique was applied prior to the measurement (32, 33). As the frequency shift is obtained in NC-AFM instead of force, we employed a conventional conversion method, where the short-range contribution in the frequency curve is extracted by subtracting the contribution of long-range van der Waals force in the form of $\Delta f \propto z^{-1.5}$ (34). The short-range part is transformed to force using the formula in ref. (35).

**Computational details.**

The DFT calculations within a generalized gradient approximation (GGA) (36, 37) were performed for the geometry optimizations using the OpenMX code (38) based on norm-conserving pseudopotentials generated with multireference energies (39) and optimized pseudoatomic basis functions (40). For each C atom, two, two, and one optimized radial functions were allocated for the *s*, *p*, and *d* orbitals. For Si and H atoms used for an AFM tip model and hydrogen terminations, s2p2d1 and s1p1 basis functions were adopted, respectively. A cutoff radius of 6 bohr was chosen for the basis functions of C and H atoms, whereas 7 bohr was chosen for Si atoms. The qualities of basis functions and fully relativistic pseudopotentials were carefully benchmarked by the delta gauge method (41) to ensure accuracy of our calculations. An electronic temperature of 300 K is used to count the number of electrons by the Fermi-Dirac function. The regular mesh of 220 Ry in real space was used for the numerical integration and for the solution of the Poisson equation (42). The geometry optimizations were performed by a combination scheme of the rational function (RF) method (43) and the direct inversion iterative sub-space (DIIS) method (44) with a BFGS update (45-48) for the approximate Hessian. The threshold of the forces for geometry optimizations was set to be 0.0003 Hartree/bohr.

In order to simulate the interaction between the tip and the surface, a five-layer $(8 \times 8)$ carbon diamond (001) surface models and an optimized Si(111) tip with the periodic boundary condition were used. The size of the surface was chosen so that the tips under the periodicity do not interfere with each other. The bottom two layers of carbon atoms of the substrate and the upper one layer of Si and H atoms of Si(111) tip are fixed during the geometry optimization. Models using different numbers of layers and different angles of the tips are discussed in Fig. S7. So, a vacuum layer greater than 25 Å was inserted between the layers to isolate their interactions.

See Ref. (28) for a computational model of the closest Wannier function.

**Acknowledgments**

The authors acknowledge K. Iwata. The computation in this work has been done using the facilities of the Supercomputer Center, the Institute for Solid State Physics, the University of Tokyo. R.Z. acknowledges the support of the WINGS-MERIT program at the University of Tokyo.

**Funding:**

This work was supported by Grant-in-Aid JSPS KAKENHI Grant Nos. 22H04496, 20H05849, 21K18867, 22H05448 and 22H01950, by JST FOREST Program (Grant Number JPMJFR203J, Japan), by the Asahi Glass Foundation, and by the Murata Science Foundation.

**Author contributions:**

Conceptualization: YS, TO, MO

Sample growth: MO, TM, DT

AFM observation: RZ, YY, YS

DFT calculation: MF, TO

Writing—original draft: RZ, YY

Writing—review & editing: RZ, YY, MF, TO, MO, TM, DT, YS

**Competing interests:** The authors declare that they have no competing interests.

**Data and materials availability:** All data are available in the main text or the supplementary materials.

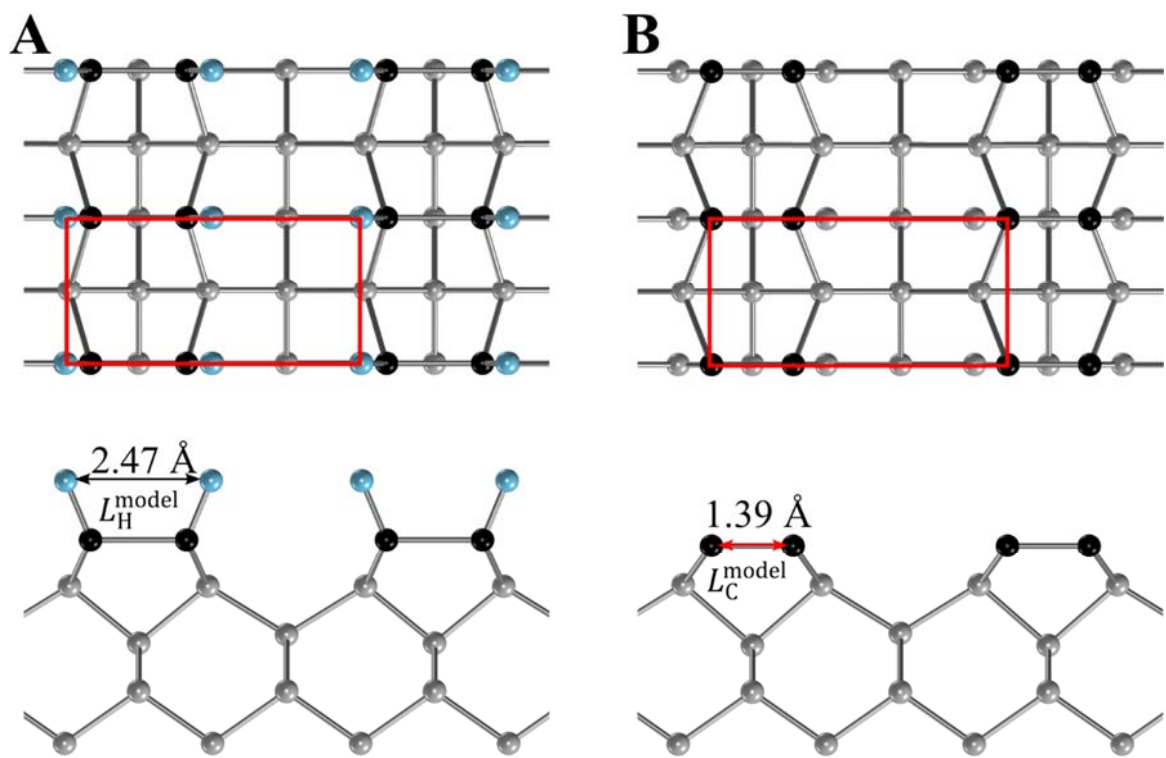

**Fig. 1. Models of hydrogenated diamond (001) surface and clean diamond (001) surface.**
Models of (A) hydrogenated diamond (001) and (B) clean diamond (001) were obtained with the DFT calculations. The top (bottom) images depict the top views (side views). The black, grey, and blue balls represent surface carbon, bulk carbon, and hydrogen atoms, respectively. The red rectangle represents the $(2 \times 1)$ reconstructed unit cells.

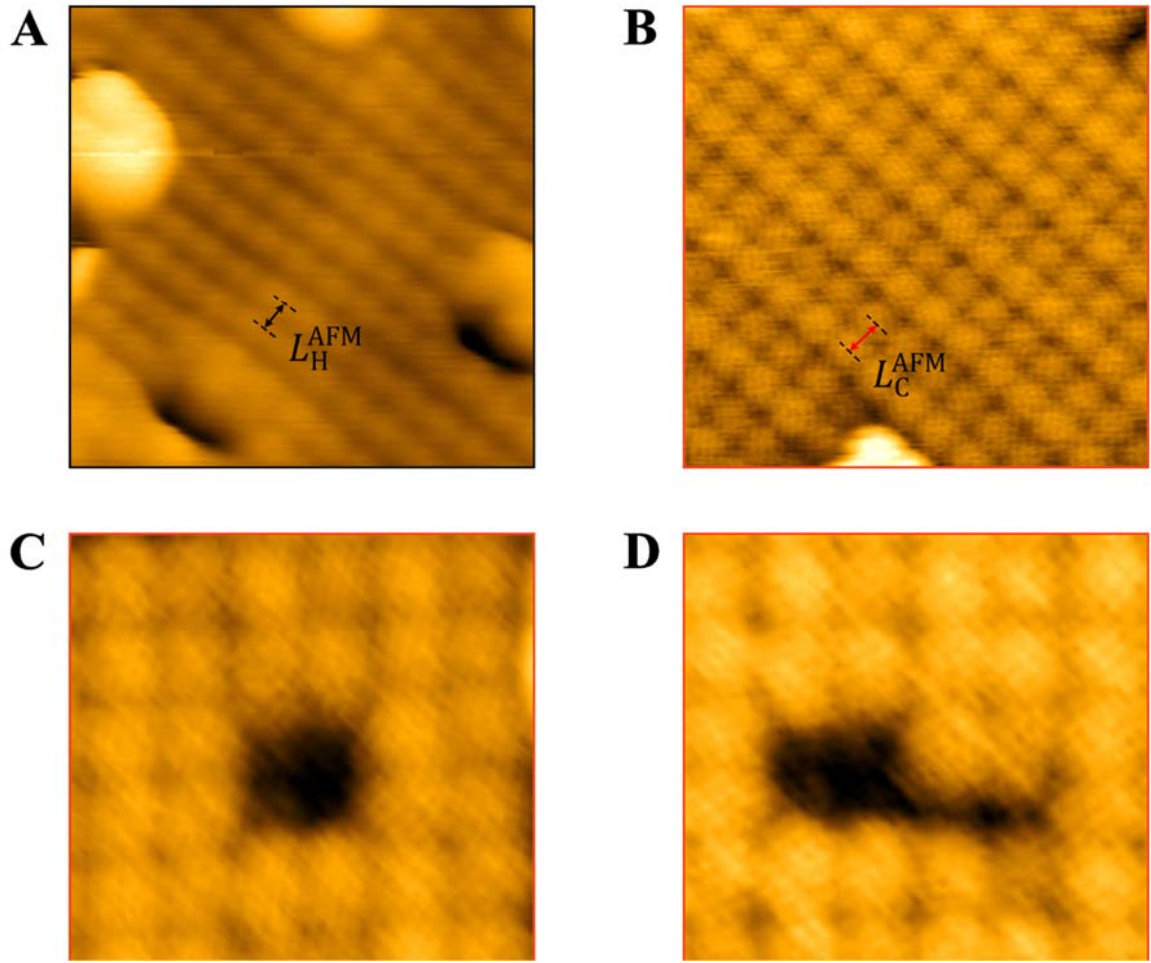

**Fig. 2. Atomic-resolution NC-AFM topographic images.** (A) An NC-AFM image for hydrogenated diamond (001). (B) A NC-AFM image with single-atom resolution for a clean diamond (001). (C) 2 dimer vacancy and (D) 3 dimer vacancy on clean diamond (001). All images were acquired at room temperature. The images size of (A) and (B) is $28 \text{ Å} \times 28 \text{ Å}$, and that of (C) and (D) is $15 \text{ Å} \times 15 \text{ Å}$. The setpoint frequency shift $\Delta f_s$, oscillation amplitude $A$, and sample bias voltage $V_s$ are (A) $\Delta f_s = -5.2$ Hz, $A = 192$ Å, $V_s = -0.55$ V and (B, C, D) $\Delta f_s = -24$ Hz, $A = 177$ Å, $V_s = 1$ V.

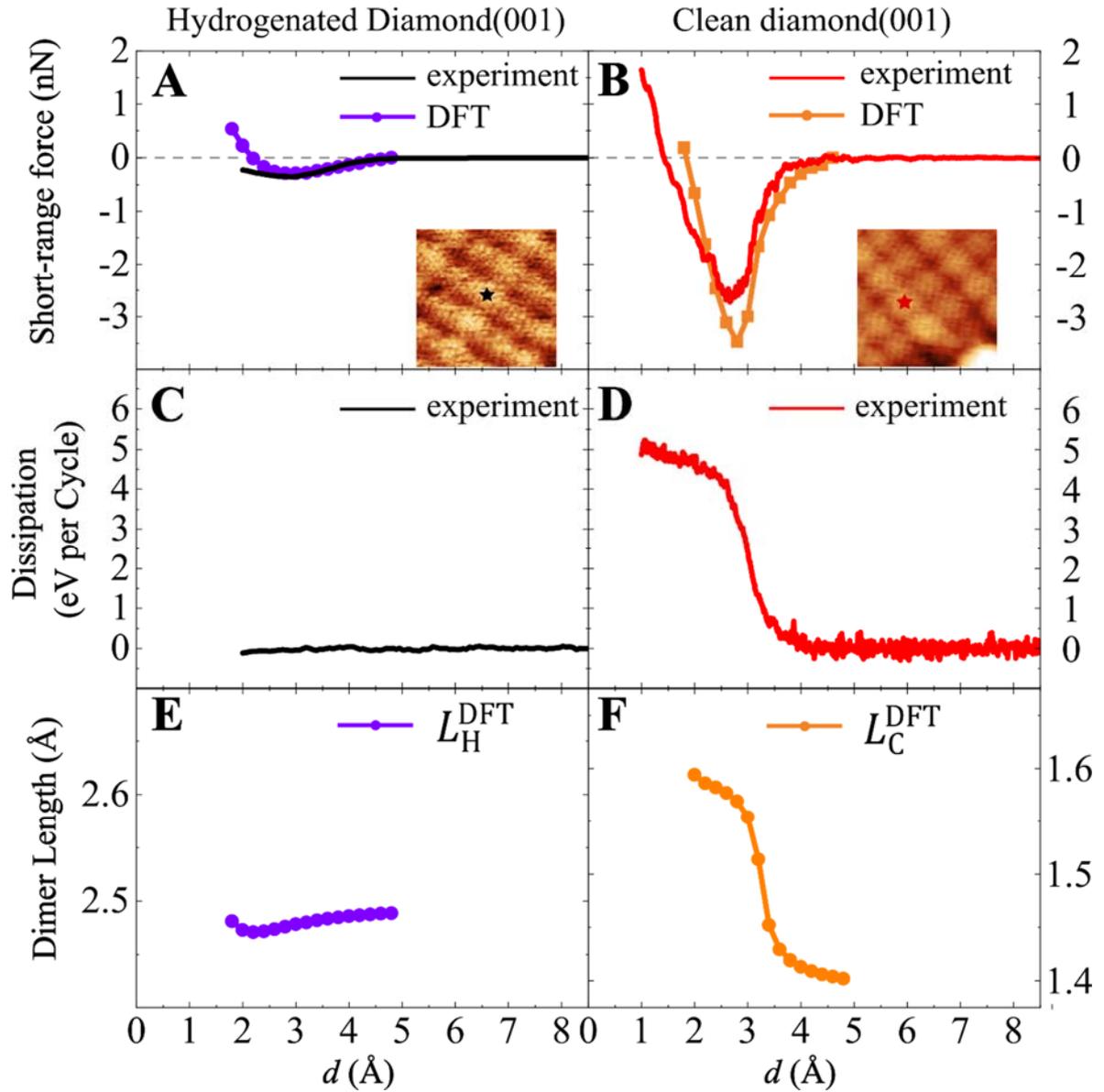

**Fig. 3. Bond formation between the diamond surface atoms and the Si tip.** Force curves for (A) the hydrogenated diamond (001) surface and (B) the clean diamond (001) surface as a function of the tip-surface distance $d$. The curves are measured at positions indicated with star marks in the NC-AFM images in insets. The dots represent the force curves obtained with the DFT calculations. Energy dissipation curves on (C) the hydrogenated diamond (001) and (D) the clean diamond (001) surfaces, which were obtained simultaneously as those in (A) and (B). Change in (E) $L_H^{DFT}$ and (F) $L_C^{DFT}$ as the tip approaches the surface. The experimental curves are shifted such that the positions of maximum attractive forces match those obtained by the calculations. For (A) and (C), $\Delta f_s = -7.5$ Hz, $A = 192$ Å, $V_s = -0.55$ V, 10 Å × 10 Å. For (B) and (D), $\Delta f_s = -24$ Hz, $A = 177$ Å, $V_s = 1$ V, 10 Å × 10 Å.

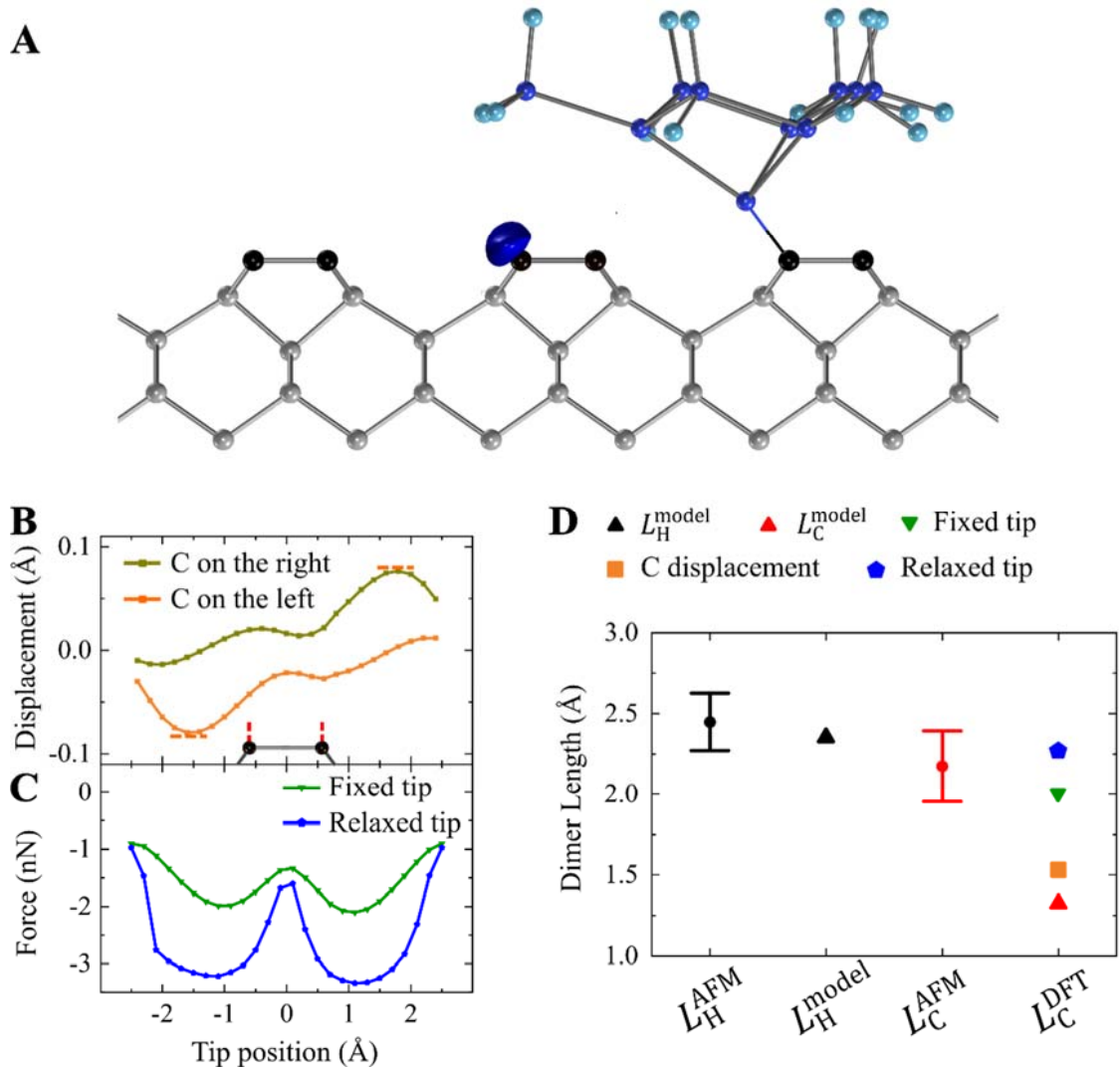

**Fig. 4. Mechanism for high resolution NC-AFM imaging**. (A) The structure used for the DFT calculation, where a silicon (111) tip scans at 3 Å above the dimers of a clean diamond (001) surface. The black, blue, and cyan balls represent dimerized carbon, silicon, and hydrogen atoms, respectively. The blue hemisphere depicts the distribution of the CWF. (B) Lateral displacement of carbon atoms as a function of the lateral tip position. The origin of the horizontal axis is chosen at the center of the dimer. (C) Force acting on the tip with and without atomic relaxation of the tip. (D) Dimer length obtained with different assumptions. Average values of the dimer lengths statistically obtained from multiple NC-AFM images are shown as circles with a standard deviation in the error bar. $L_C^{model}$ is represented in red triangle, and $L_C^{DFT}$ considering the displacement of C atoms in orange square. Distance between the local minima in force for a fixed/relaxed tip is in green inverted triangle/blue pentagon.

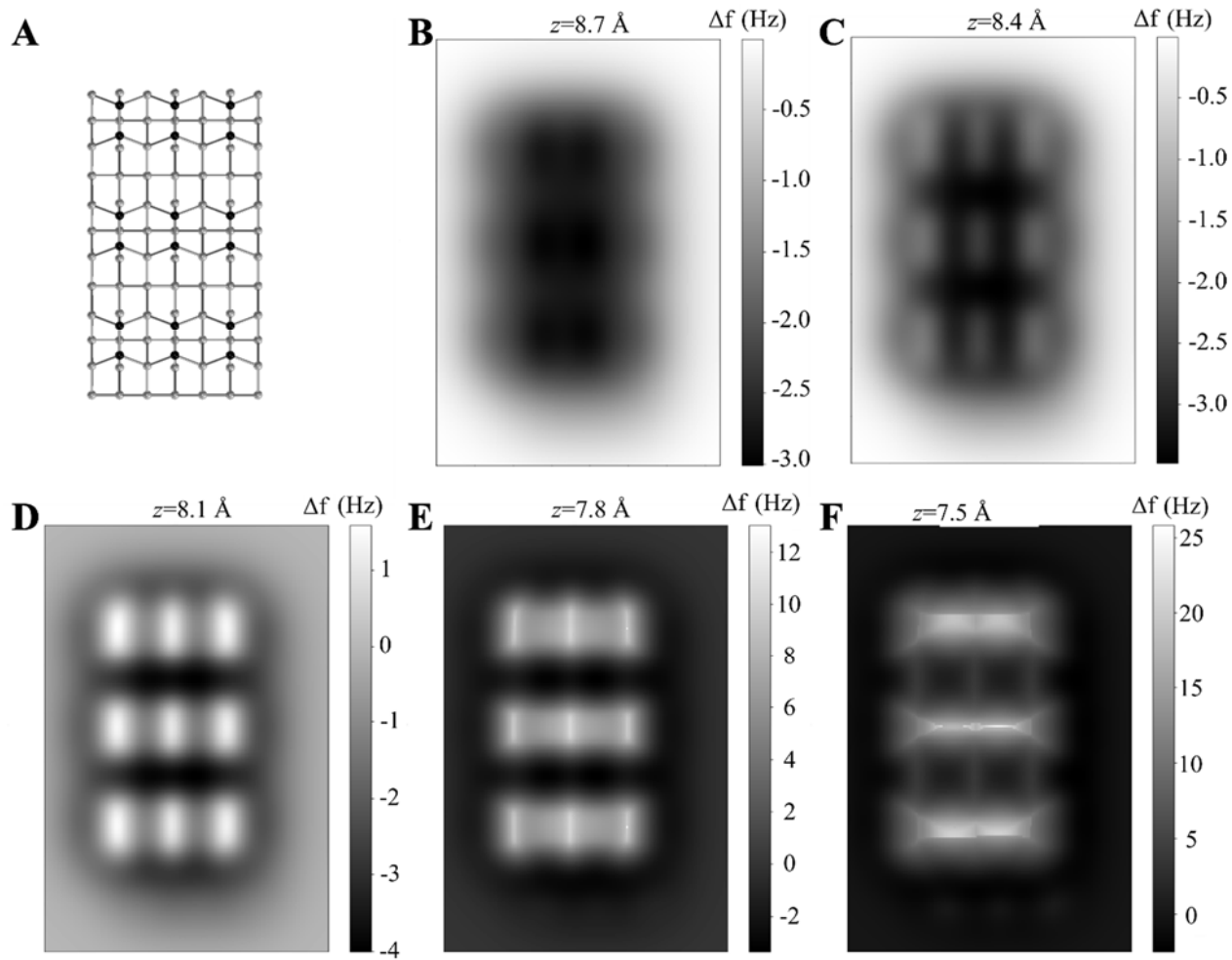

**Fig. S1. Simulated AFM images using a CO tip. (A)** Model of a clean diamond (001) surface used for the simulation. **(B)-(F)** Frequency-shift images of constant height scans at different tip height. The probe particle model was used (49).

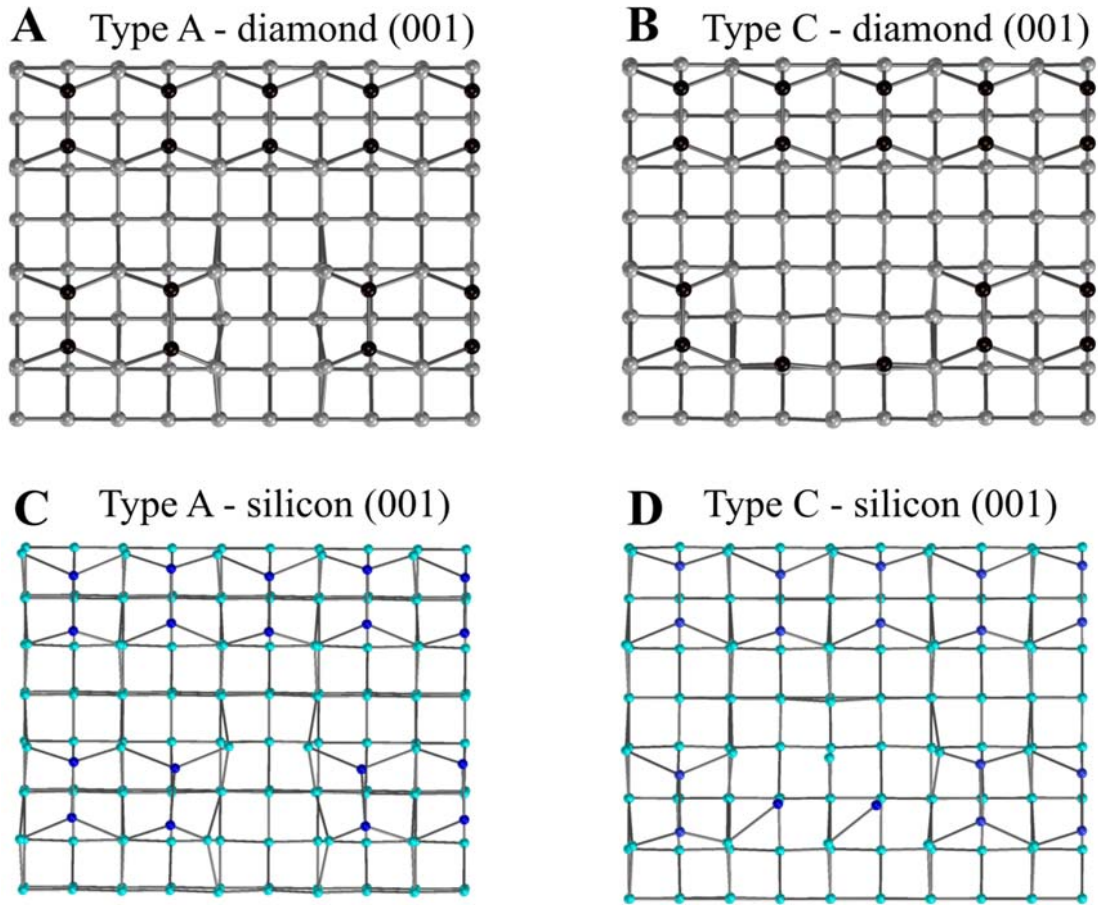

**Fig. S2. DFT calculation of defects on diamond (001).** **(A)** Model of the Type A defect for diamond (001). The formation energy is 9.45 eV. **(B)** Model of the Type C defect for diamond (001). The formation energy is 15.18 eV. **(C)** Model of Type A defect for silicon (001). The formation energy is 7.27 eV. **(D)** Model of Type C defect for silicon (001). The formation energy is 8.47 eV. The large difference in the formation energy between Type A and Type C for diamond (001) indicates that Type C is virtually unavailable. The formation energy is defined as $E_{formation} = E_{\text{free carbon (silicon) dimer}} + E_{\text{surface with hole}} - E_{\text{surface}}$. Here, the definition of $E_{\text{surface}}$ is the energy of diamond (silicon) (001) surface. $E_{\text{free carbon (silicon) dimer}}$ represents the energy of an individual dimer. $E_{\text{surface with hole}}$ signifies the energy of the surface with hole. A smaller formation energy suggests greater stability, indicating a higher probability of formation.

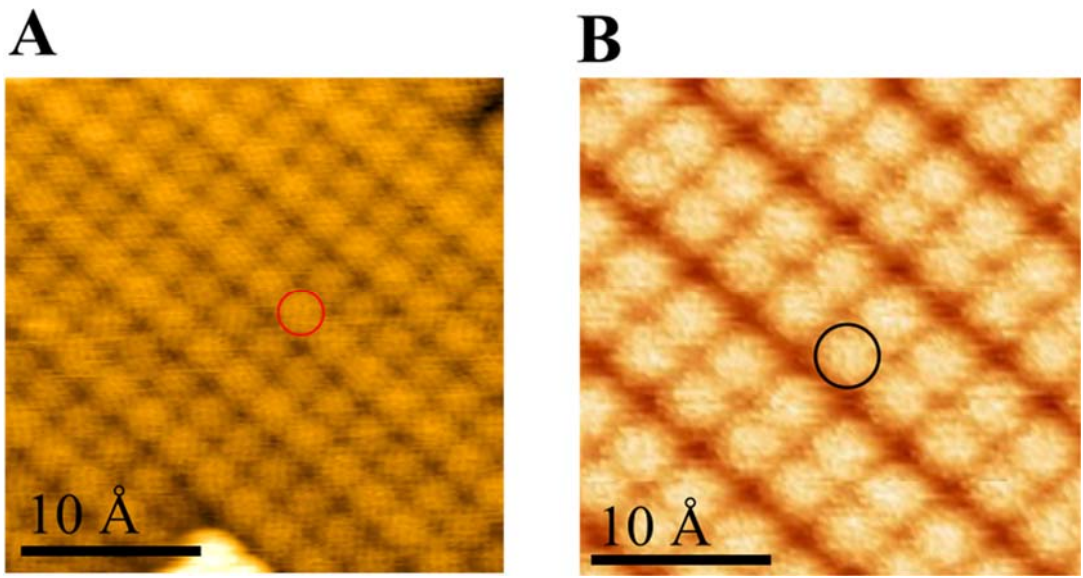

**Fig. S3. AFM images of C(001) and Si(001).** **(A)** AFM image of C(001). The diameter of carbon atoms is approximately 2.6 Å (red circle). $f_0 = 167$ kHz, $\Delta f_s = -24$ Hz, $A = 177$ Å, $V_s = 1$ V. **(B)** AFM images of Si(001). The diameter of silicon atoms is approximately 3.7 Å (black circle). $f_0 = 178$ kHz, $\Delta f_s = -28.9$ Hz, $A = 261$ Å, $V_s = -0.33$ V.

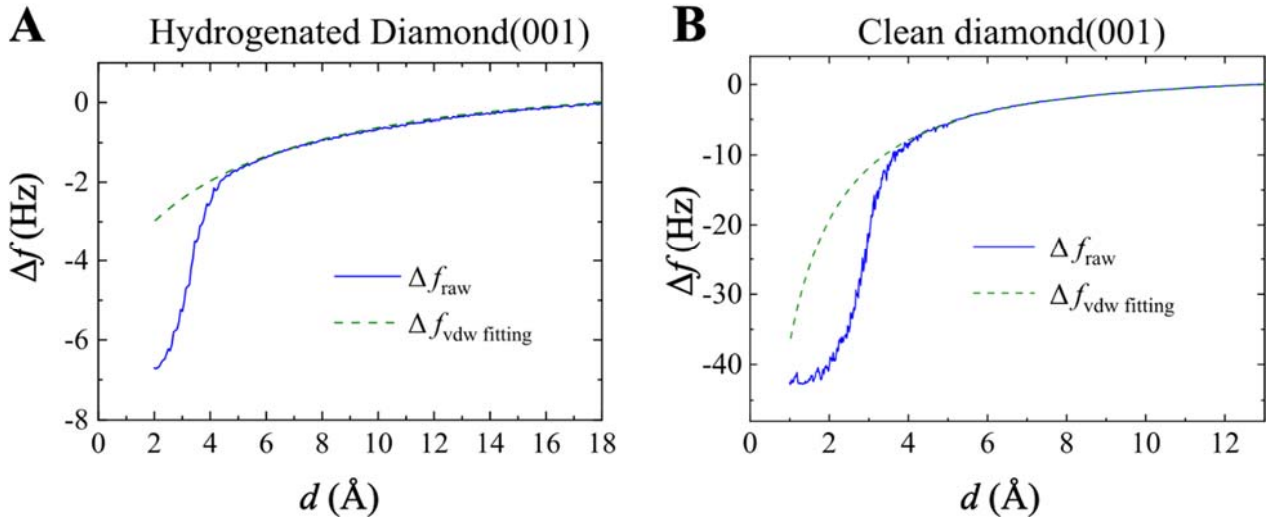

**Fig. S4. Raw frequency-shift curves for Figs. 3(A) and 3(B).** Frequency shift as a function of tip-sample distance for **(A)** hydrogenated diamond (001) and **(B)** clean diamond (001). The blue solid lines represent the raw data before the subtraction of long-range components. The green dashed lines represent the fitting results for the long-range van der Waals interactions. The fitting range is 5 Å to the maximum distances.

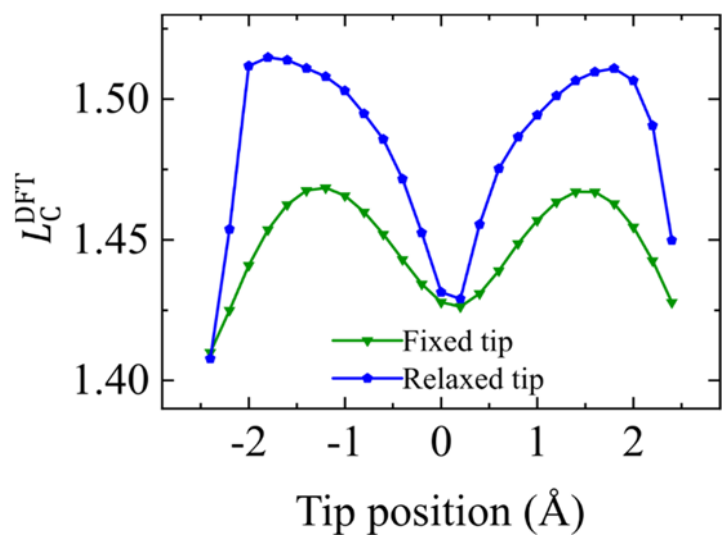

**Fig. S5. Dimer length as the function of lateral tip position.** The actual dimer length changes with the lateral position of the tip. A relaxed tip enhances the displacement of C atoms due to stronger attractive force.

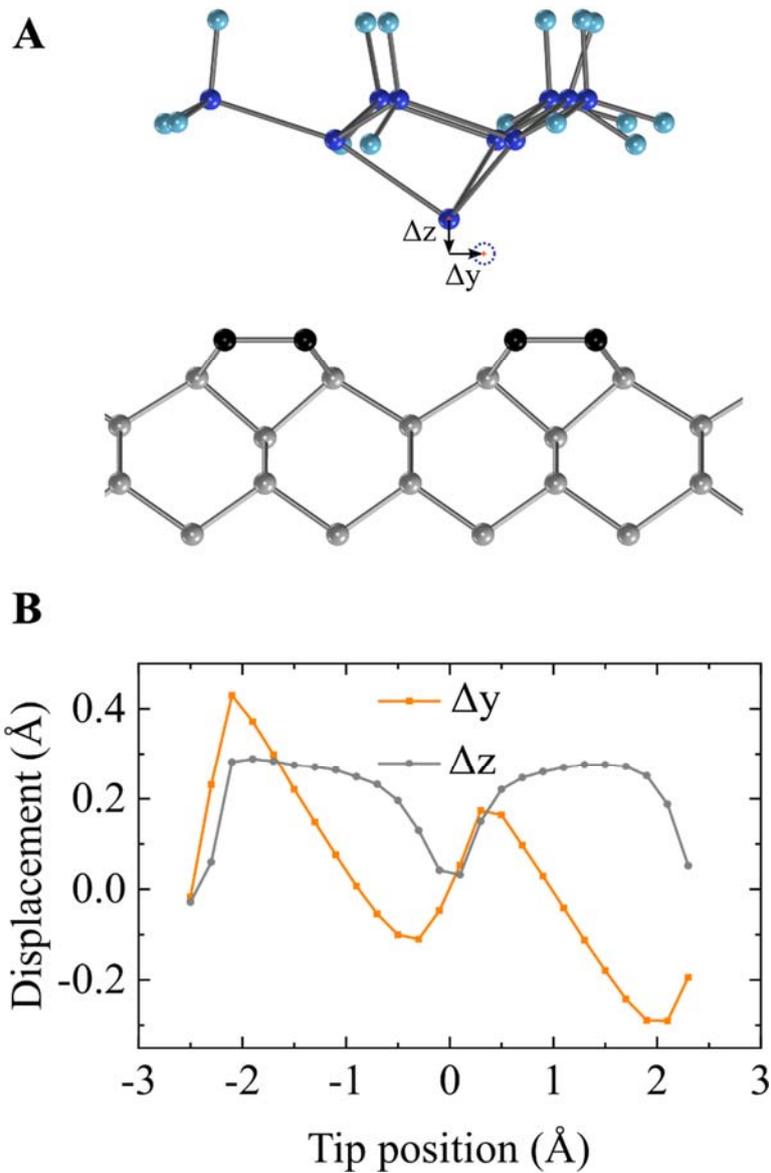

**Fig. S6. Displacement of silicon atoms at the tip apex. (A)** Model of a Si(111) tip and a clean diamond (001) surface. The position of the tip apex Si atom after relaxation is depicted with a dashed circle. **(B)** Displacement of the tip apex Si atom. The origin for the tip position is chosen at the center between the dimer.

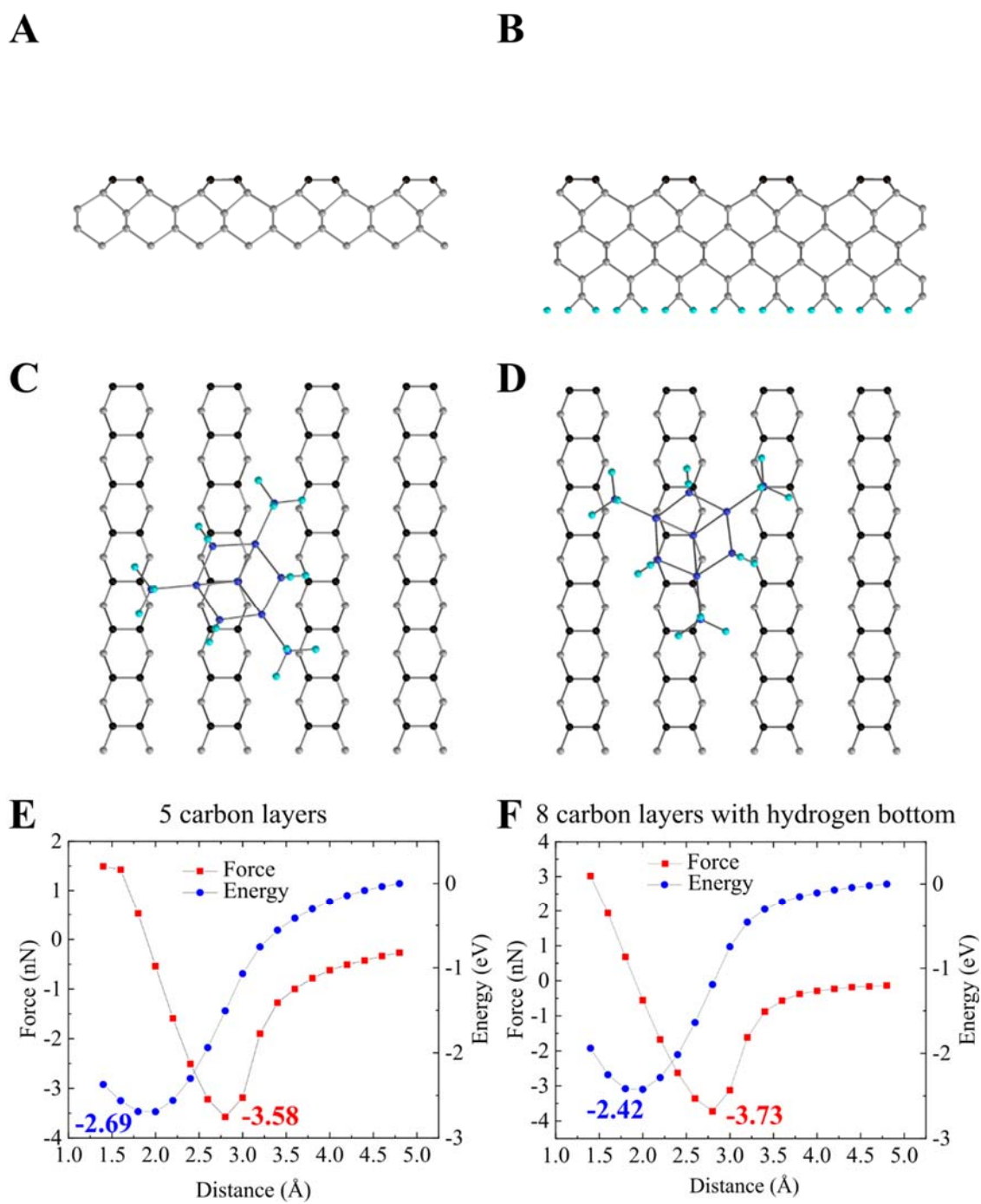

**Fig. S7. Force curves for different number of layers and different orientation of tips. (A), (B)** Side view of the models with different thickness. **(C), (D)** Top view of the models with different tip orientation. **(E)** Force and energy curves for (A) and (C). **(F)** Force and energy curves for (B) and (D).